\documentclass[prl,twocolumn,showpacs,amsmath,amssymb,superscriptaddress]{revtex4-1}

\usepackage{graphicx}
\usepackage{epsfig}
\usepackage{color}
\usepackage{epsf}
\usepackage{dcolumn}
\usepackage{textcomp}
\usepackage{amsmath,amstext,amssymb}
\usepackage{enumitem}
\usepackage{hyperref}
\usepackage[ansinew]{inputenc}
\usepackage{bm}
\usepackage{braket}
\usepackage{siunitx}
\usepackage{mnsymbol}

\PassOptionsToPackage{numbers,sort&compress}{natbib}

\makeatletter
\g@addto@macro\normalsize{%
  \setlength\abovedisplayskip{4pt}
  \setlength\belowdisplayskip{4pt}
  \setlength\abovedisplayshortskip{4pt}
  \setlength\belowdisplayshortskip{4pt}
}
\makeatother

\newcommand{\be}{\begin{equation}}
\newcommand{\ee}{\end{equation}}
\newcommand{\bea}{\begin{eqnarray}}
\newcommand{\eea}{\end{eqnarray}}

\begin{document}

\title{A spin heat engine coupled to a harmonic-oscillator flywheel}

\author{D.~von Lindenfels}
\affiliation{QUANTUM, Institut f\"ur Physik, Universit\"at Mainz, Staudingerweg 7, 55128 Mainz, Germany}
\author{O.~Gr\"ab}
\affiliation{QUANTUM, Institut f\"ur Physik, Universit\"at Mainz, Staudingerweg 7, 55128 Mainz, Germany}
\author{C.~T.~Schmiegelow}\thanks{Present address: LIAF -  Laboratorio de Iones y Atomos Frios, Departamento de Fisica \& Instituto de Fisica de Buenos Aires, 1428 Buenos Aires, Argentina}
\affiliation{QUANTUM, Institut f\"ur Physik, Universit\"at Mainz, Staudingerweg 7, 55128 Mainz, Germany}
\author{V.~Kaushal}
\affiliation{QUANTUM, Institut f\"ur Physik, Universit\"at Mainz, Staudingerweg 7, 55128 Mainz, Germany}
\author{J.~Schulz}
\affiliation{QUANTUM, Institut f\"ur Physik, Universit\"at Mainz, Staudingerweg 7, 55128 Mainz, Germany}
\author{Mark T.~Mitchison}
\affiliation{School of Physics, Trinity College Dublin, College Green, Dublin 2, Ireland}
\author{John Goold}
\affiliation{School of Physics, Trinity College Dublin, College Green, Dublin 2, Ireland}

\author{F.~Schmidt-Kaler}
\affiliation{QUANTUM, Institut f\"ur Physik, Universit\"at Mainz, Staudingerweg 7, 55128 Mainz, Germany}
\author{U.~G.~Poschinger}\email{poschin@uni-mainz.de}
\affiliation{QUANTUM, Institut f\"ur Physik, Universit\"at Mainz, Staudingerweg 7, 55128 Mainz, Germany}

\begin{abstract}
We realize a heat engine using a single electron spin as a working medium. The spin pertains to the valence electron of a trapped $^{40}$Ca$^+$ ion, and heat reservoirs are emulated by controlling the spin polarization via optical pumping. The engine is coupled to the ion's harmonic-oscillator degree of freedom via spin-dependent optical forces. The oscillator stores the work produced by the heat engine and therefore acts as a flywheel. We characterize the state of the flywheel by reconstructing the Husimi $\mathcal{Q}$ function of the oscillator after different engine runtimes. This allows us to infer both the deposited energy and the corresponding fluctuations throughout the onset of operation, starting in the oscillator ground state. In order to understand the energetics of the flywheel, we determine its ergotropy, i.e. the maximum amount of work which can be further extracted from it. Our results demonstrate how the intrinsic fluctuations of a microscopic heat engine fundamentally limit performance.
\end{abstract}

\maketitle

Heat engines converting thermal energy to mechanical work have always been the centerpiece of thermodynamics. They consist of four fundamental components: a working agent, the cold and hot heat reservoirs, and a mechanism for deposition or extraction of the generated work. Recently, thermal machines have been experimentally demonstrated in the microscopic regime \cite{ste11,bli12,mar16} and are currently entering the realm of well-controlled atomic systems: A single-ion heat engine \cite{ros16} and an ion-crystal based refrigerator \cite{Maslennikov2019} have been demonstrated recently, and engines based on ensembles of NV centers in diamond \cite{Klatzow2019}, superconducting circuits \cite{Koski2014} or ensembles of nuclear spins in a NMR setup \cite{pet18} have been studied. With decreasing size of the constituent parts and at finite operation timescales, well-established notions such as \textit{work}, \textit{heat} and \textit{efficiency} need to be reassessed \cite{cur75,ali79,talkner2016}. In particular, far from the thermodynamic limit, fluctuations play a central role \cite{jarzynski1997,sei12,AN2014}. For engines comprising a few microscopic degrees of freedom, the impact of quantum effects has been subject to theoretical studies \cite{kos84,scu02,bru14,Campo2014,Campisi2016}. 

\begin{figure}[h!tp]\begin{center}
\includegraphics[width=\columnwidth, trim={1,7cm 4,2cm 0,8cm 2,4cm},clip]{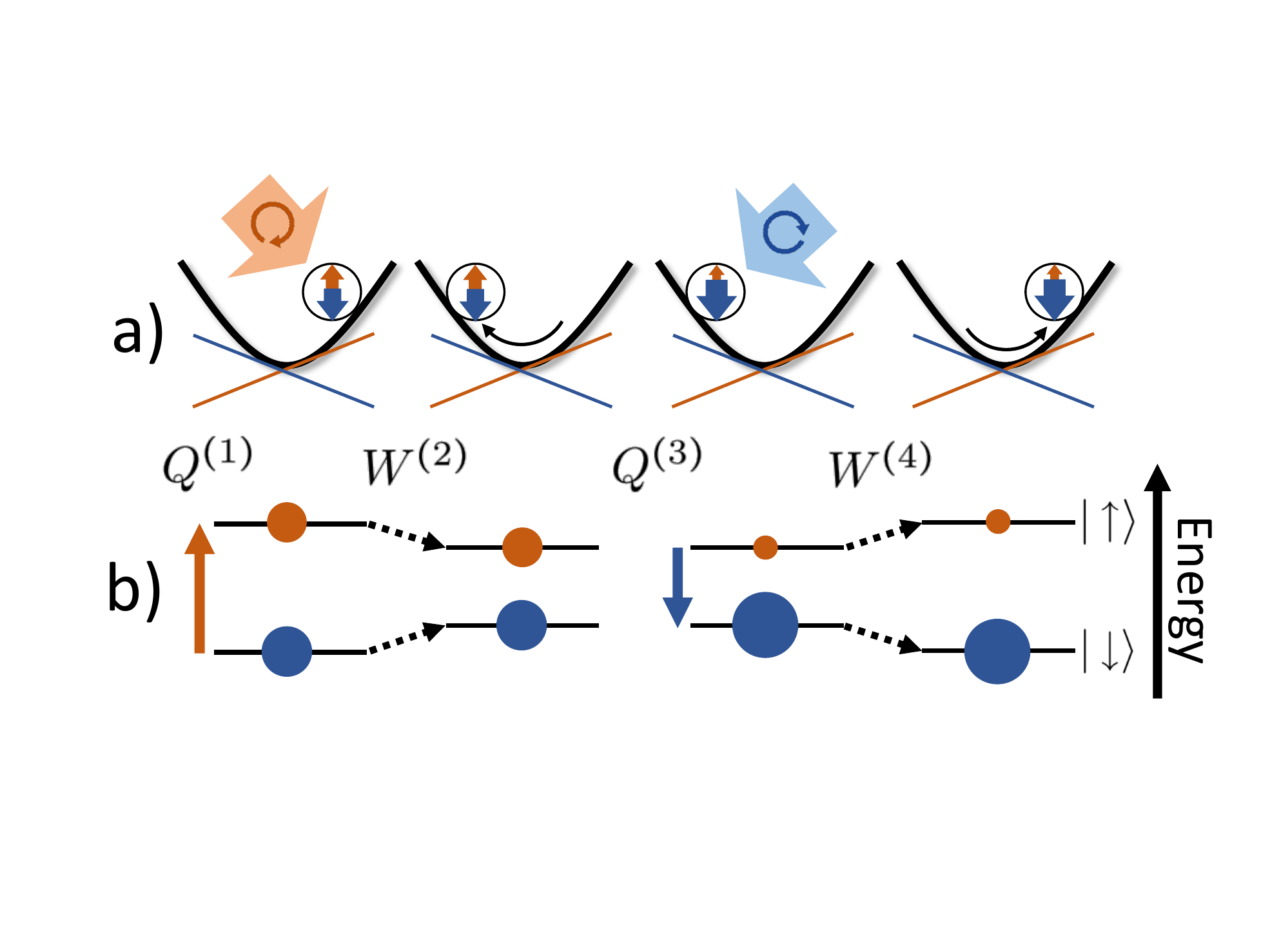}
\caption{ Operation of the four-stroke engine. \textbf{a)} Mechanical picture: The parabolas show the harmonic trap potential and lines indicate the additional spin-dependent optical potential acting on $\ket{\uparrow}$ (red) and $\ket{\downarrow}$ (blue). The arrows within the circles representing the ion correspondingly indicate the spin populations. \textbf{b)} Energy representation: The levels indicate the Zeeman energies of $\ket{\uparrow}$ and $\ket{\downarrow}$, and the size of the circles indicates the populations. Shown are the states of the system after each of the engine strokes, from left to right: isochoric heating, isentropic expansion, isochoric cooling and isentropic compression (see text). }
 \label{fig:fig1}
\end{center}\end{figure}

Here, we report on the experimental realization of a heat engine based on a two-level system as a working agent, which is coupled to a harmonic-oscillator degree of freedom \cite{Watanabe2017}, where output energy is deposited throughout the operation of the engine. It is henceforth referred to as the \textit{flywheel} \cite{LEVY2016}. The engine and flywheel degrees of freedom both allow for direct control. This enables the characterization of the energy deposition throughout the onset of the engine operation, at an energy resolution below the single quantum level. Starting with the flywheel initialized in the ground state, we characterize its state after different engine operation times by reconstructing its Husimi $\mathcal{Q}$ function \cite{LV2017}. From this, we infer the energy deposited in the flywheel along with its fluctuations. The measured fluctuations have a significant thermal component, indicating that not all of the energy transferred to the flywheel is extractable work. Therefore, in order to quantify the work done by the engine we evaluate the ergotropy~\cite{ALLAHVERDYAN2004,Gelbwaser2013,GhoshPNAS}, i.e., an upper bound on the amount of  work which can be extracted from the flywheel. 
The results reveal how the generation of useful work is limited by effects which are characteristic for microscopic systems.

\textit{Engine operation.---}The heat engine operates on the spin of the valence electron pertaining to a single trapped $^{40}$Ca$^+$ ion. The operation is depicted in Fig.~\ref{fig:fig1}. Heating and cooling of the spin is achieved by controlling its polarization in an external magnetic field via alternating optical pumping. The harmonic motion of the ion in the confining Paul trap acts as the flywheel. We place the ion in an optical standing wave (SW), which mediates the coupling between the engine and flywheel via a spin-dependent optical dipole force \cite{POSCHINGER2010,SCHMIEGELOW2016} along the oscillation ($x$) direction. The trap center $x=0$ coincides with a node of the SW. The Hamiltonian of the coupled spin-oscillator system reads
\begin{equation}
\hat{H} = \hat{H}_{\rm HO} + \hbar\left(\omega_z + \Delta_S\sin(k_\mathrm{SW}\hat{x}) \right) \frac{\hat{\sigma}_z}{2},
	\label{eq:hamiltonian}
\end{equation}
where $\omega_z$ denotes the Zeeman splitting of the spin and $\hat{\sigma}_z$ is the Pauli $z$ operator. The bare flywheel Hamiltonian is $\hat{H}_{\rm HO}=\hbar \omega_t\left(\hat{n}+\tfrac{1}{2}\right)$, where $\omega_t$ is the trap frequency along $x$ and $\hat{n}$ is the number operator. The parameter {\small $\Delta_S$} denotes the amplitude of the SW in terms of the spatially varying ac-Stark shift, where $k_\mathrm{SW}\approx 2\pi/$~280~nm is the effective wavenumber. The internal energy is given by the Zeeman energy of the spin: $U = \hbar \omega'_z(\langle\hat{x}\rangle)\langle\hat{\sigma}_z\rangle/2$. For small displacements $k_\mathrm{SW}\langle\hat{x}\rangle \ll1$, the effective Zeeman shift --- the sum of the magnetic field-induced shift and ac Stark shift from the SW --- is $\omega'_z(\langle\hat{x}\rangle) = \omega_z+${\small $\Delta_S$}$ k_\mathrm{SW}\langle\hat{x}\rangle$.

\begin{figure}[htp]\begin{center}
\includegraphics[width=\columnwidth,trim={0 16.5cm 0 0},clip]{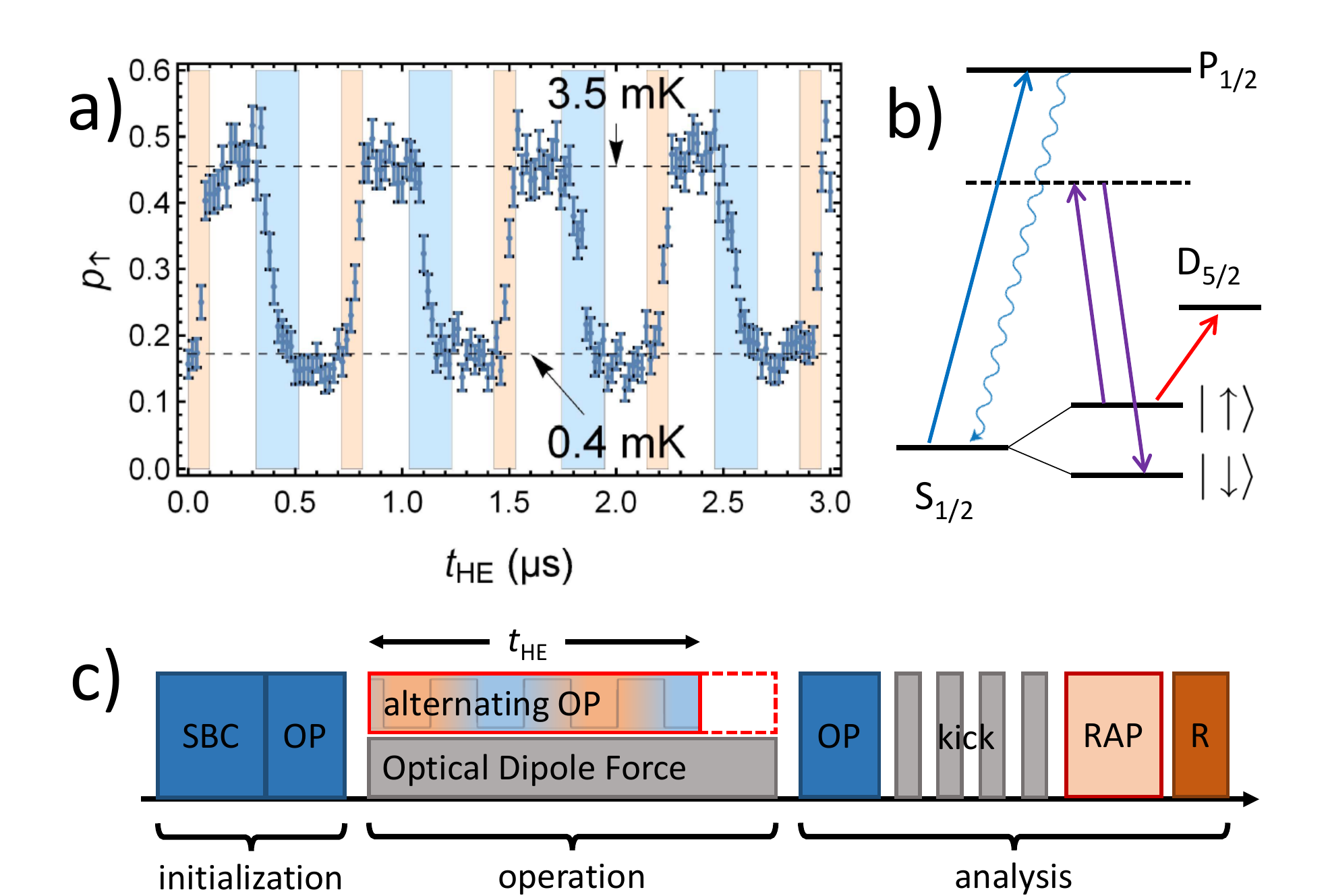}
\caption{\textbf{a)} Measured probabilities to find the spin in $\ket{\uparrow}$ throughout the engine operation. The colored areas indicate that the pump laser is switched on (pink: heating, blue: cooling). The equilibrium probabilities indicated by the horizontal dashed lines indicate the optical pumping operations, emulating the equilibration with reservoirs at temperatures $T_C$ and $T_H$. \textbf{b)} Relevant atomic levels of $^{40}$Ca$^+$, showing the working-medium levels $\ket{\downarrow}$ and $\ket{\uparrow}$, the transition to the metastable $D_{5/2}$ level utilized for spin readout (red arrow), the stimulated Raman transition for probing (purple arrows) and the cycling transition utilized for optical pumping and readout (blue arrows). \textbf{c)} Experimental sequence for the reconstruction of the flywheel $\mathcal{Q}$ function (see text), indicating sideband cooling (SBC), optical pumping (OP), rapid adiabatic passage (RAP) and spin readout (R).}
\label{fig:fig2}
\end{center}\end{figure}

Optical pumping with optical polarization alternating at the trap period $2\pi/\omega_t$ emulates the coupling to reservoirs: After each pumping step, the populations of the Zeeman sublevels of the S$_{1/2}$ electronic ground state correspond to a fixed temperature, see Fig.~\ref{fig:fig2}~a). The cold reservoir temperature $T_C$ corresponds to predominant population of the lower-energy Zeeman sublevel, i.e. $\langle\hat{\sigma}_z\rangle \gtrsim -1$, while the hot reservoir temperature $T_H>T_C$ corresponds to predominant depolarization, $\langle\hat{\sigma}_z\rangle \lesssim 0$. The hot and cold temperatures are determined via 
\begin{equation}
\langle\hat{\sigma}_z\rangle=-\tanh({\hbar \omega'_z}/{2 k_B T}). 
\label{eq:tempfrompop}
\end{equation}

Close to the the SW node, the ion experiences a mean spin-dependent force $\mathcal{F}=-\hbar k_{\text{SW}}\Delta_S\langle\hat{\sigma}_z\rangle/2$. Since $\langle\hat{\sigma}_z\rangle$ varies periodically at frequency $\omega_t$, this leads to an average resonant driving force on the oscillator, i.e.\ deposition of work in the flywheel. The engine is equivalent to a four-stroke Otto motor: Associating the effective Zeeman shift $\omega_z'$ with the inverse volume of a working gas in a macroscopic engine, we identify the four strokes of the cycle as follows, see Fig.~\ref{fig:fig1}: The first optical pumping step realizes isochoric heating of the spin (heat transfer $Q^{(1)}$). For an ion positioned at $x>0$, the effective restoring force is increased. In the second step, the harmonic oscillation half-cycle leads to a decrease of $\omega_z'$, i.e. isentropic expansion (consumption of work $W^{(2)}$ from the flywheel), as the ion moves to $x<0$. Isochoric cooling takes place in the third step (heat transfer $Q^{(3)}$). This step again increases the effective restoring force. Then, the final oscillation half-cycle leads to an increase of $\omega_z'$, i.e. isentropic compression (release of work $W^{(4)}$ to the flywheel). As energy is continuously stored in the flywheel, the amplitude of the harmonic oscillation increases during the operation of the engine. Since the internal Zeeman energy of the spin scales with the oscillator displacement, the cycle is not closed, and the power increases with the number of cycles.

\textit{Quantifying work.---}Due to its coupling with the baths, the spin's orientation is intrinsically uncertain, giving rise to a random spin-dependent force acting on the flywheel. This leads to fluctuations in the energy transferred to the flywheel during the isentropic strokes. Even for an ideal Otto cycle with fast, perfectly timed isochores and disregarding other experimental imperfections, the flywheel executes a random walk in phase space, whose statistical properties are determined by the equilibrium spin populations~\cite{SUPPLEMENTAL}. As a result, only a fraction of the deposited energy constitutes useful, extractable work, while the remainder increases the flywheel’s entropy.

The flywheel's work content is quantified by its \textit{ergotropy}, i.e. the maximum work that can be extracted via a cyclic unitary transformation~\cite{ALLAHVERDYAN2004}. It is defined as $\mathcal{W} = {\rm Tr} [\hat{H}_{\rm HO}\hat{\rho}] -  {\rm Tr} [\hat{H}_{\rm HO}\hat{\rho}_p]$, where $\hat{\rho}$ is the state of the flywheel and $\hat{\rho}_p$ is the \textit{passive state} unitarily related to $\hat{\rho}$~\cite{SUPPLEMENTAL}. The ergotropy represents the amount of ordered energy stored in the flywheel while disregarding random contributions such as thermal fluctuations. Measuring the engine's work output thus requires us to characterize the state of the flywheel resulting from operation of the engine.

\textit{Experimental realization.---}We store a single $^{40}$Ca$^+$ ion trapped in a miniaturized Paul trap \cite{SCHULZ2008}, at a secular trap frequency of $\omega_t\approx2\pi\times\SI{1.4}{\mega\hertz}$ along the $x$-axis. The Zeeman sublevels of the $S_{1/2}$ electronic ground state, i.e. the two-level system working agent of the engine, are denoted by $\ket{\uparrow}$ and $\ket{\downarrow}$ (Fig.~\ref{fig:fig2}~b). A constant magnetic field yields a Zeeman splitting between these of $\omega_z \approx 2\pi\times\SI{13}{\mega\hertz}$. The alternating optical pumping is carried out via laser pulses driving the S$_{1/2}\leftrightarrow$P$_{1/2}$ cycling transition near 397~nm, at pulse durations shorter than half the trap period $\pi/\omega_t$. For the hot (cold) isochore, the optical polarization is dynamically set to left (right) circular by means of an electro-optical modulator, which leads to population transfer $\ket{\downarrow}\rightarrow \ket{\uparrow}$ ($\ket{\uparrow}\rightarrow \ket{\downarrow}$). The intensities and pulse durations determine the spin polarizations at the end of the isochores and therefore the effective bath temperatures. We work with equilibrium spin polarizations of $\langle\hat{\sigma}_z\rangle^{(H)}=$~-0.084(4) and $\langle\hat{\sigma}_z\rangle^{(C)}=$-0.656(6), which correspond to temperatures $T_H=\SI{3.5\pm.2}{\milli\kelvin}$ and $T_C=\SI{0.40\pm.01}{\milli\kelvin}$ according to Eq.~\eqref{eq:tempfrompop}. The SW - providing the coupling between spin and flywheel - is generated by two laser beams far-detuned from the cycling transition and controlled via acousto-optical modulators. This gives rise to a spin-dependent ac Stark shift, periodically varying along $x$ at an amplitude of {\small $\Delta_S$}$=2\pi\times\SI{2.73\pm.02}{\mega\hertz}\ll \omega_z$. 

The experimental sequence is depicted in Fig.~\ref{fig:fig2}~c). In each experimental run, the flywheel is initialized in its ground state via resolved sideband cooling~\cite{POSCHINGER2009}, and the spin is initialized to a statistical mixture state corresponding to temperature $T_C$ via optical pumping. Then, the SW is switched on we run the heat engine for a time $t_\textrm{HE}$, during which the alternating pumping is carried out. 

\begin{figure}[t!p]\begin{center}
\includegraphics[width=0.8\columnwidth, trim={0cm 0cm 0cm 2cm},clip]{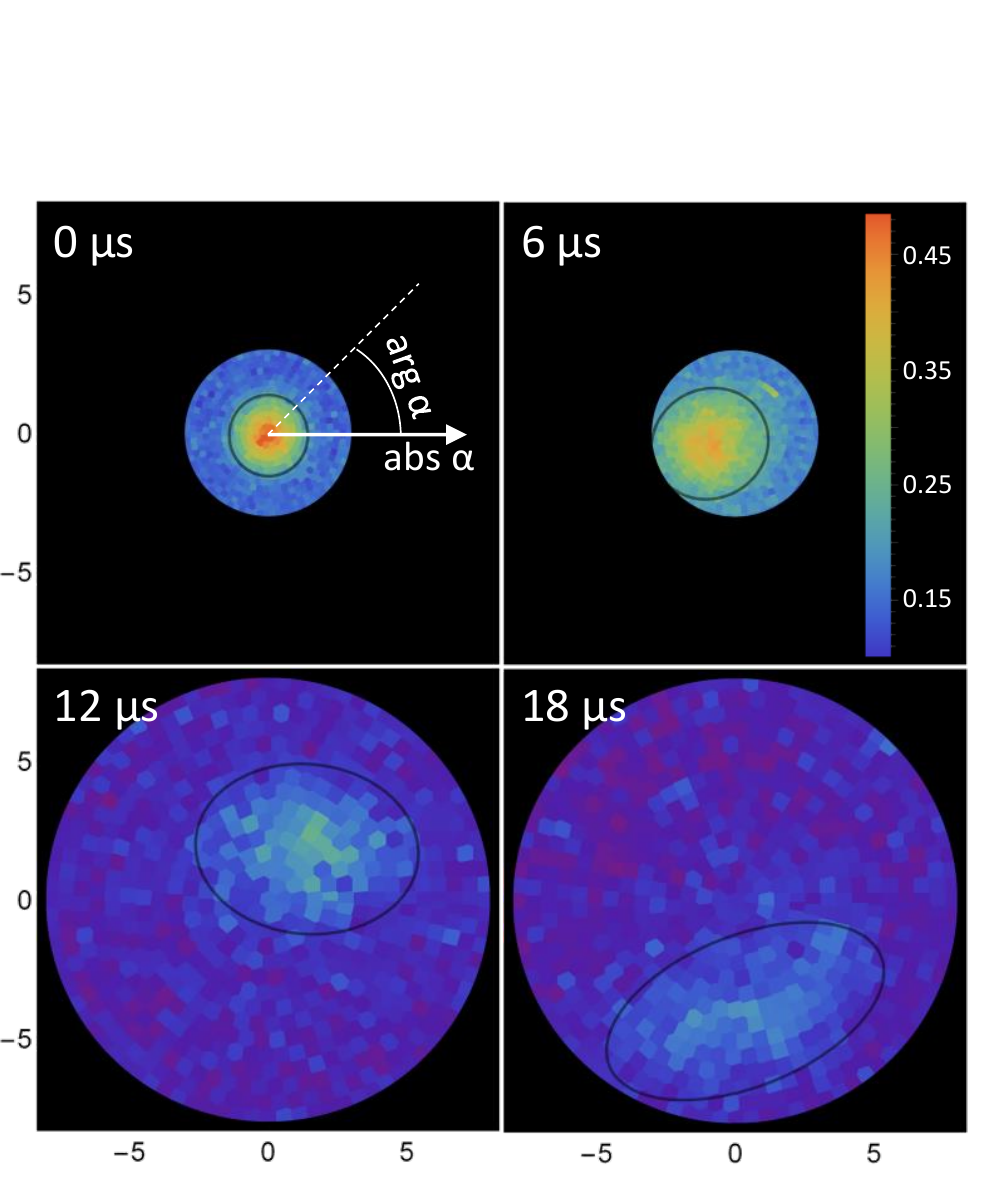}
\caption{Measured $\mathcal{Q}$ functions (raw data) for the flywheel at different times throughout the heat engine operation. Each pixel shows the result of 1000 independent experimental runs, and corresponds to a kick voltage determining $\vert\alpha\vert$ and a kick delay determining the phase $\arg\alpha$. The black lines are $1/e^2$ contours pertaining to fits of the $\mathcal{Q}$ function to the model Eq.~\eqref{eq:DSTmodel}. $\vert\alpha\vert = 1$ corresponds to an oscillation amplitude of \SI{19}{\nano\meter}. For further evaluation, the raw data values are shifted and rescaled to account for imperfect population transfer and readout, such that the normalization $\int \mathcal{Q}(\alpha,\alpha^*) \mathrm{d}^2\alpha=1$ is fulfilled, and that $\mathcal{Q}(\alpha,\alpha^*)$ assumes zero for large values of $\vert\alpha\vert$.}
 \label{fig:qfunction}
\end{center}\end{figure}

After heat engine operation throughout $t_{\rm HE}$, the SW is switched off. Then, the spin is pumped to $\ket{\downarrow}$ and its role is changed --- rather than driving the engine, it is now employed as a probe for the final state of the flywheel $\hat{\rho}$. As the flywheel was initialized close to its ground state and energies in the few-quanta regime are to be resolved, a quantum-mechanical measurement scheme is ultimately required. We reconstruct the $\mathcal{Q}$ function of the flywheel 
\begin{equation}
\mathcal{Q}(\alpha,\alpha^*)=\tfrac{1}{\pi}\bra{0}\hat{D}^{\dagger}(\alpha)\hat{\rho}\hat{D}(\alpha)\ket{0}.
\label{eq:qfunctiondef}
\end{equation}
This quantity is the probability to find the flywheel in the ground state after application of a displacement kick $\hat{D}(\alpha)$, and represents a quasi-probability distribution in phase space. The state reconstruction measurement starts with a displacement 'kick' operation of complex amplitude $\alpha$ on the flywheel. This operation is carried out by applying calibrated voltage pulses to neighboring trap segments \cite{ZIESEL2013}. After the kick, the population of all states $\ket{n,\downarrow}$ is transferred to $\ket{n-1,\uparrow}$. This is possible \textit{only} for $n\neq 0$, therefore only the population pertaining to $n=0$ remains in $\ket{\downarrow}$. This is realized via rapid adiabatic passage (RAP) on the first red sideband of the stimulated Raman transition between $\ket{\uparrow}$ and $\ket{\downarrow}$. Finally, spin readout via population transfer $\ket{\uparrow}\rightarrow\text{D}_{5/2}$ to a metastable state \cite{POSCHINGER2009} and subsequent detection of state-dependent fluorescence upon driving the cycling transition yields a 'bright' result at a probability corresponding to the $\mathcal{Q}$ function value Eq.~\eqref{eq:qfunctiondef}. A similar method has been used e.g. in Refs.~\cite{AN2014,LV2017}.

The $\mathcal{Q}$ function is reconstructed in polar phase space coordinates by scanning $\vert\alpha\vert$ via the kick voltage amplitude and $\arg\alpha$ via the kick delay time with respect to the onset of the heat engine operation. For increasing values of $\vert\alpha\vert$, the resolution of $\arg\alpha$ is increased, such that the support of $\mathcal{Q}(\alpha,\alpha^*)$ in phase space is scanned at roughly constant steps. 

\textit{Results.---}We reconstruct $\mathcal{Q}(\alpha,\alpha^*)$ for different heat engine runtimes $t_{\textrm{HE}}$, in steps of $t_{\textrm{HE}}^{(i)}=i\;\Delta t_{\textrm{HE}}$ with $\Delta t_{\textrm{HE}} = \SI{3}{\micro\second}$, up to a duration of about 25 flywheel oscillation periods. Examples of reconstructed $\mathcal{Q}$ functions are shown in Fig.~\ref{fig:qfunction}, revealing the nature of the final flywheel states. The quasi-probability peaks around a fixed amplitude and phase, indicating coherent oscillations. Furthermore, the support of the distribution increases asymmetrically beyond the uncertainty limit, indicating a thermal component induced by spin fluctuations and squeezing by the anharmonic SW potential. We therefore model the resulting flywheel states as displaced squeezed thermal states (DSTS):
\begin{eqnarray}
\label{eq:DSTmodel}
\hat{\rho}_{\rm DST}(\beta,\zeta,\bar{n})&=&\hat{D}(\beta)\hat{S}(\zeta)\hat{\rho}_{\rm th}(\bar{n})\hat{S}^{\dagger}(\zeta)\hat{D}^{\dagger}(\beta), \\
\hat{\rho}_{\rm th}(\bar{n}) &=& \sum_n \frac{\bar{n}^n}{(\bar{n}+1)^{n+1}} \ket{n}\bra{n},
\label{eq:thermal_state}
\end{eqnarray}
with the thermal state $\hat{\rho}_{\rm th}(\bar{n})$ pertaining to the mean thermal phonon number $\bar{n}$, the squeezing operator $\hat{S}(\zeta)$ and the displacement operator $\hat{D}(\beta)$. The squeezing excitation is small as compared to thermal and displacement excitations. For obtaining estimates of the parameters $\bar{n},\beta,\zeta$ for each reconstructed flywheel state, we fit the model Eq.~\eqref{eq:DSTmodel} to given $\mathcal{Q}$ function data. To that end, for each test parameter set $\{\beta,\zeta,\bar{n}\}$, a density matrix is computed in a truncated number state basis from Eq.~\eqref{eq:DSTmodel}, from which the $\mathcal{Q}$ function values at the probed phase space coordinates are computed directly from Eq.~\eqref{eq:qfunctiondef}. The fit minimizes the root-mean-square difference between the measured and model $\mathcal{Q}$ function values.

The DSTS model provides a description of the flywheel energetics. The ergotropy $\mathcal{W}$ and mean energy $E = {\rm Tr}[\hat{H}_{\rm HO} \hat{\rho}_{\rm DST}]$ are given respectively by~\cite{SUPPLEMENTAL} 
\begin{align}
\label{eq:ergotropy_explicit}
\mathcal{W} &= \hbar \omega_t|\beta|^2 + \hbar \omega_t\sinh^2(|\zeta|)(2\bar{n}+1),\\
\label{eq:energy_explicit}
E &= \mathcal{W} + \hbar\omega_t\bar{n}.
\end{align}
The dominant contribution to the ergotropy derives from the oscillatory motion represented by $\beta$, with a further squeezing contribution. Conversely, thermal fluctuations increase the mean energy by an amount $\hbar\omega_t \bar{n}$, that cannot be extracted as work. Note, however, that squeezing catalyzes the extraction of work from thermal fluctuations~\cite{GhoshPNAS} via the term proportional to $\sinh^2(|\zeta|)\bar{n}$ in Eq.~\eqref{eq:ergotropy_explicit}.

\begin{figure}\begin{center}
		\includegraphics[width=0.8\linewidth]{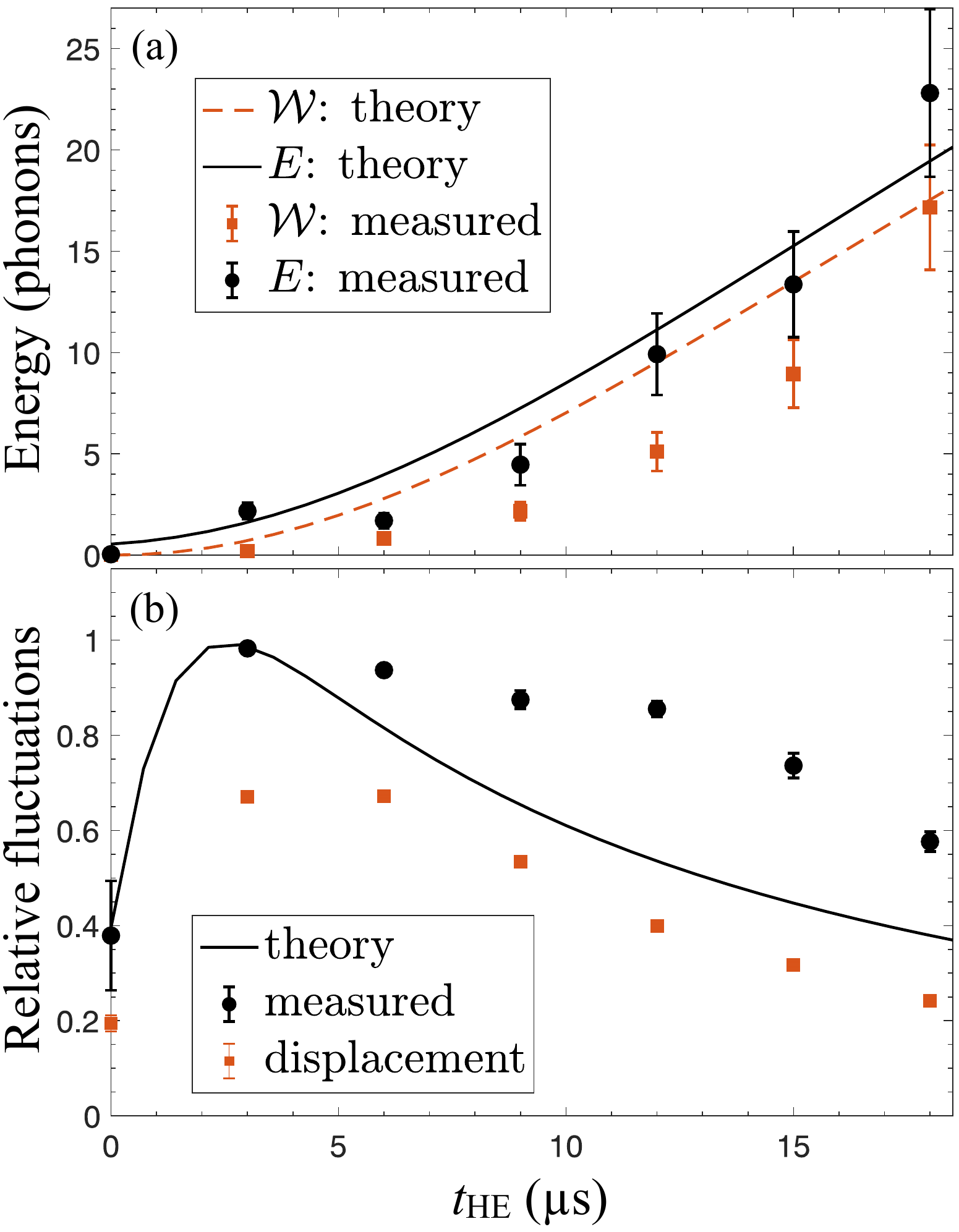}
		\caption{Results: (a) measured energy $E$, ergotropy $\mathcal{W}$ and (b) relative energy fluctuations $\Delta E/E$, compared to (cycle-averaged) predictions of the master equation~\cite{SUPPLEMENTAL}. In (b) we also plot the relative fluctuations of a coherent state with the measured displacement $\beta$, i.e., $(\Delta E/E)_{\rm displ} = |\beta|/(|\beta|^2+\tfrac{1}{2})$. In the simulations, the flywheel starts in a thermal state with the measured initial energy. Note that the relative fluctuation values exhibit small error bars as both statistical and systematic errors of $\Delta E$ and $E$ are correlated. \label{fig:results}}
\end{center}\end{figure}

The energy and ergotropy deposited in the flywheel are displayed in Fig.~\ref{fig:results}, together with the relative energy fluctuations $\Delta E/E$, where $\Delta E^2 = {\rm Tr}[\hat{H}_{\rm HO}^2\hat{\rho}_{\rm DST}]- E^2$. The experimental results show qualitative agreement with simulations of a Lindblad master equation describing the Otto cycle. Importantly, our theoretical model incorporates the full Hamiltonian~\eqref{eq:hamiltonian}, which is nonlinear in $\hat{x}$. The assumption that the ion remains close to the SW node, so that $k_{\rm SW}\langle \hat{x}\rangle \ll 1$, breaks down after about five engine cycles. As a consequence, the engine transitions from its initial onset behavior, with ergotropy increasing quadratically in time, to a later regime where the curvature of the SW potential limits the growth of ergotropy to be approximately linear. The squeezing contribution to the ergotropy amounts to 1.9(3)~quanta at $t_{\rm HE}=\SI{18}{\micro\second}$.

Our measurements show that the flywheel's ergotropy $\mathcal{W}$ remains strictly less than its energy $E$ due to the presence of thermal excitation. However, the \textit{fraction} $\mathcal{W}/E$ grows over time, indicating an increasingly ordered deposition of energy in the flywheel. This is reflected in the behavior of $\Delta E/E$, which exhibits a crossover from an initial transient increase dominated by thermal fluctuations to asymptotic decay at longer times~\cite{SUPPLEMENTAL}. Note that even a pure coherent state, which would arise from unitary transfer of work to the flywheel, would still exhibit Poissonian energy fluctuations. As shown in Fig.~\ref{fig:results}(b), the measured energy fluctuations significantly exceed this ``displacement limit''. These results demonstrate that the extractable work produced by microscopic engines is reduced by intrinsic fluctuations. However, in order to distinguish useless thermal energy from useful deposited work, one must go beyond energy statistics to quantitatively describe the thermodynamic performance of such engines --- for which ergotropy is the relevant quantity.

The obtained ergotropy values fall significantly short from the simulation, while the relative energy fluctuations exceed the simulation values. This discrepancy between theory and experiment can be attributed to imperfections such as photon recoils during optical pumping, phase jitters of the SW and off-resonant scattering from the SW, which are not included in the simulation. See the Supplemental Material for details of the theoretical model and error analysis~\cite{SUPPLEMENTAL}.

\textit{Conclusion \& outlook.---}We have experimentally demonstrated the operation of a single spin-$\tfrac{1}{2}$ heat engine coupled to a harmonic-oscillator flywheel, and we have characterized the finite-time thermodynamic performance of the combined engine-flywheel system. Furthermore, we have shown that $\mathcal{Q}$-function measurements together with a DSTS ansatz allow for an accurate assessment of the energetic capability of our microscopic engine via the ergotropy, i.e.\ the maximum amount of work which can be extracted from the flywheel by a cyclic unitary protocol.  Our results reveal the importance of fluctuations in machines operating on single atomic degrees of freedom. 

We stress that while our measurement method is intrinsically quantum mechanical, and while we initialize the flywheel in its ground state, the resulting states of the flywheel are consistent with a semi-classical model. This is a consequence of the operational principle implemented here, which requires optical pumping, i.e.strong incoherent coupling of the spin engine to reservoirs to accomplish heat transfer. 

Ultimately, one would seek to establish reservoirs consisting of sets of trapped ions rather than external control fields, which would open up a plethora of possibilities for studying thermal machines comprised of well-controlled microscopic quantum systems. Further extensions of the spin heat engine could encompass limit-cycle operation by adding persistent laser cooling of the flywheel, and demonstrating autonomous operation~\cite{Tonner2005,Gelbwaser2014}. We also note that irreversible entropy production can be inferred from $\mathcal{Q}$ functions via the Wehrl entropy~\cite{Santos2018} and that our platform may allow investigation of links between ergotropy and correlations~\cite{francica2017daemonic}. Our experiment opens the door to further explorations of nano-scale thermodynamics where a work repository is explicitly included.

\begin{acknowledgments}
We acknowledge financial support by the JGU Mainz, helpful discussions with Martin Plenio, and early-stage contributions by Marcelo Luda and Johannes Rossnagel. JG is supported by a SFI-Royal Society University Research Fellowship. JG and MTM acknowledge funding from the European Research Council (ERC) under the European Union's Horizon 2020 research and innovation program (grant agreement No. 758403). FSK and UGP acknowledge funding from DeutscheForschungsgemeinschaft (FOR 2724).\end{acknowledgments}

\bibliographystyle{apsrev4-1}
\bibliography{lit}

\begin{figure}[htp]\begin{center}
\hspace*{-1cm}
\includegraphics[width=1.1\textwidth]{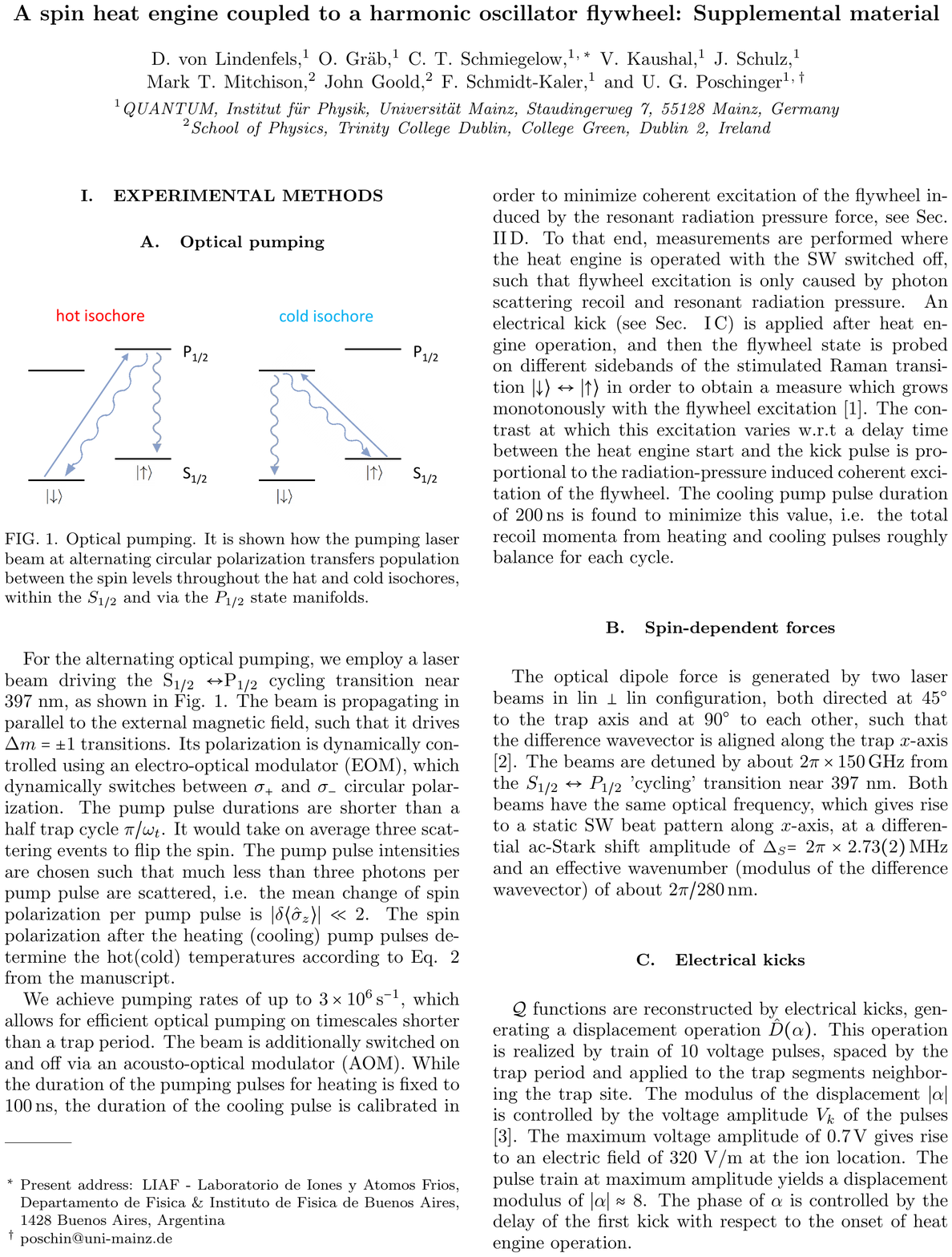}
\end{center}\end{figure}

\begin{figure}[htp]\begin{center}
\hspace*{-1cm}
\includegraphics[width=1.1\textwidth]{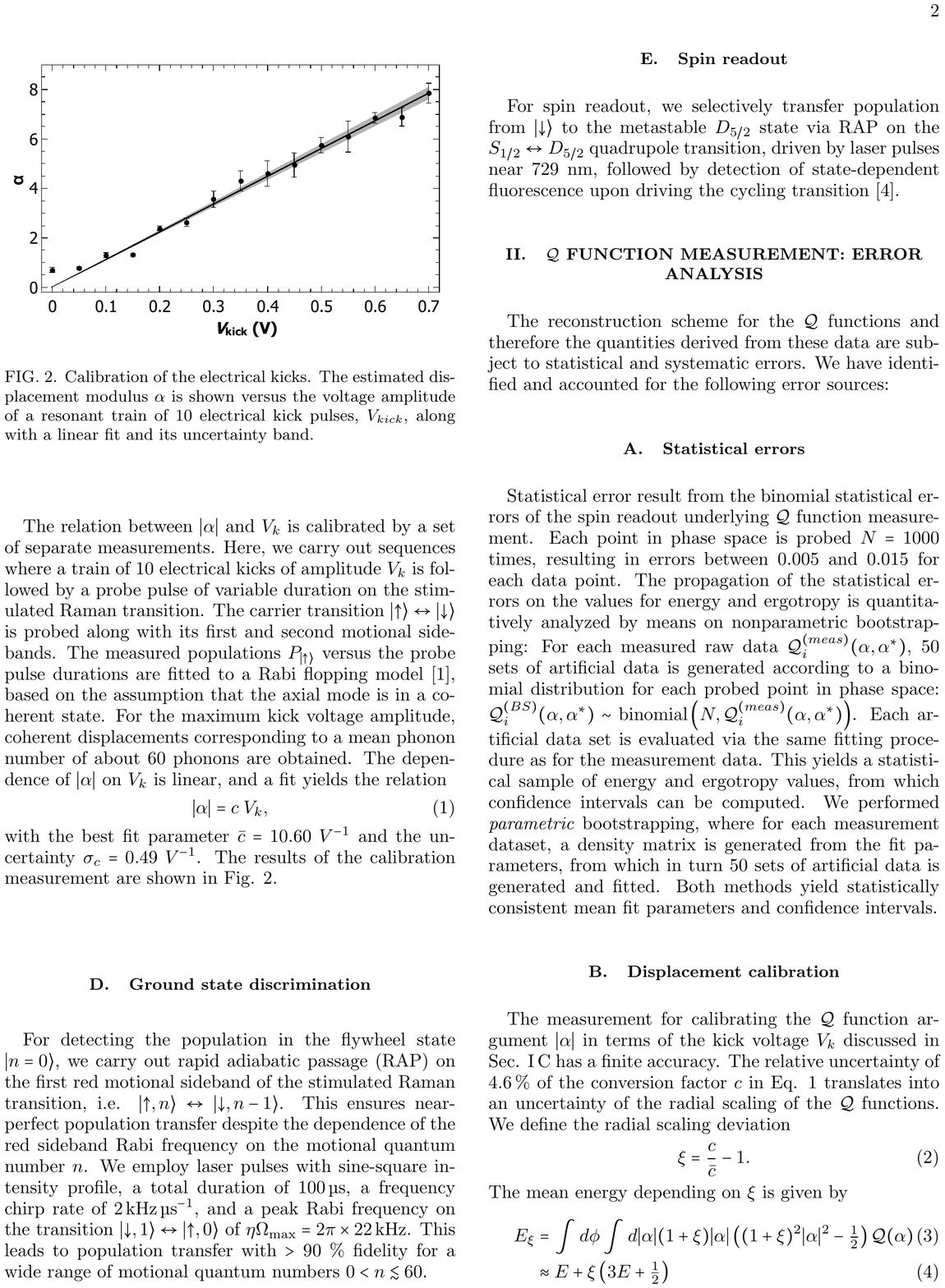}
\end{center}\end{figure}

\begin{figure}[htp]\begin{center}
\hspace*{-1cm}
\includegraphics[width=1.1\textwidth]{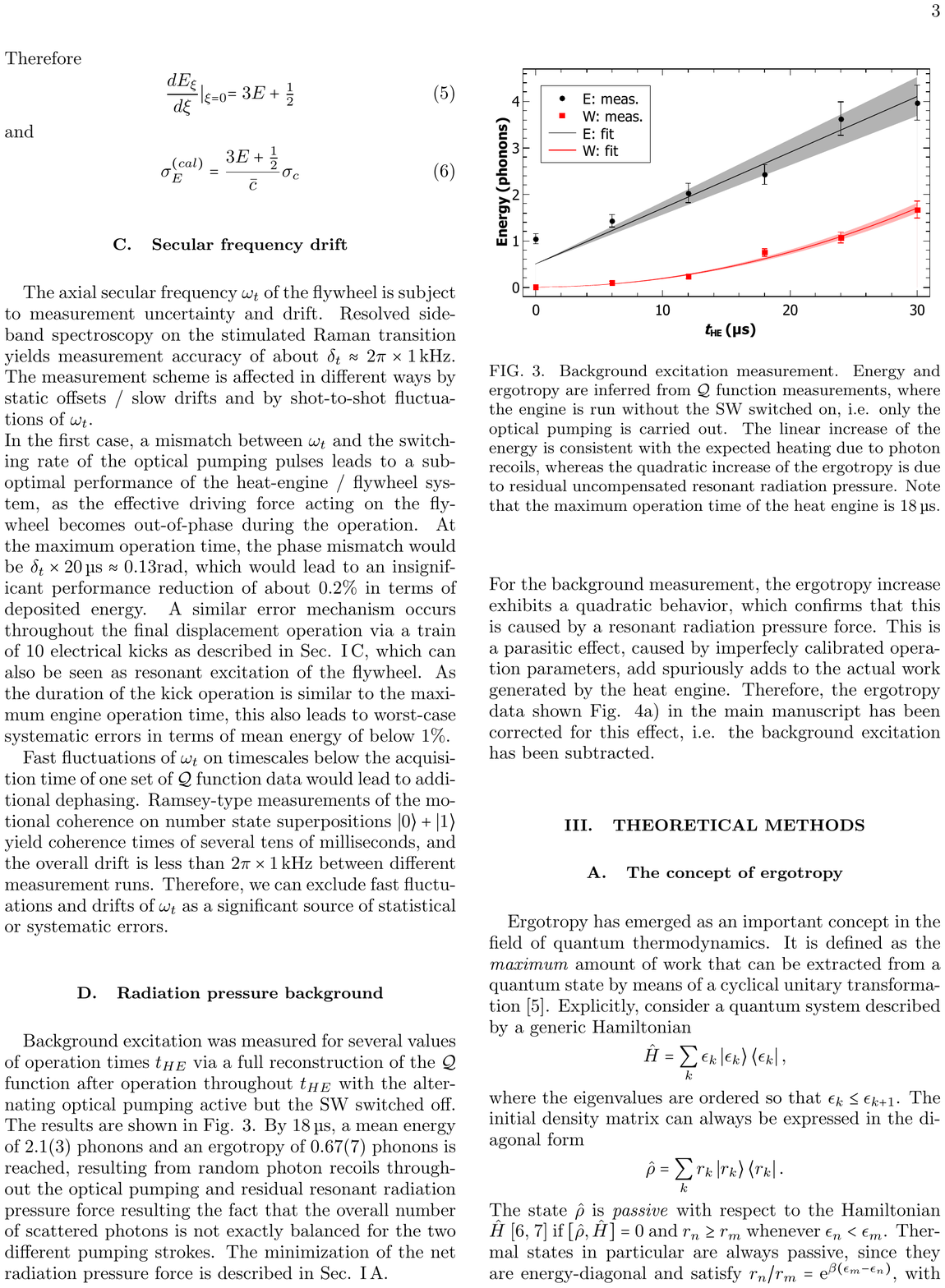}
\end{center}\end{figure}

\begin{figure}[htp]\begin{center}
\hspace*{-1cm}
\includegraphics[width=1.1\textwidth]{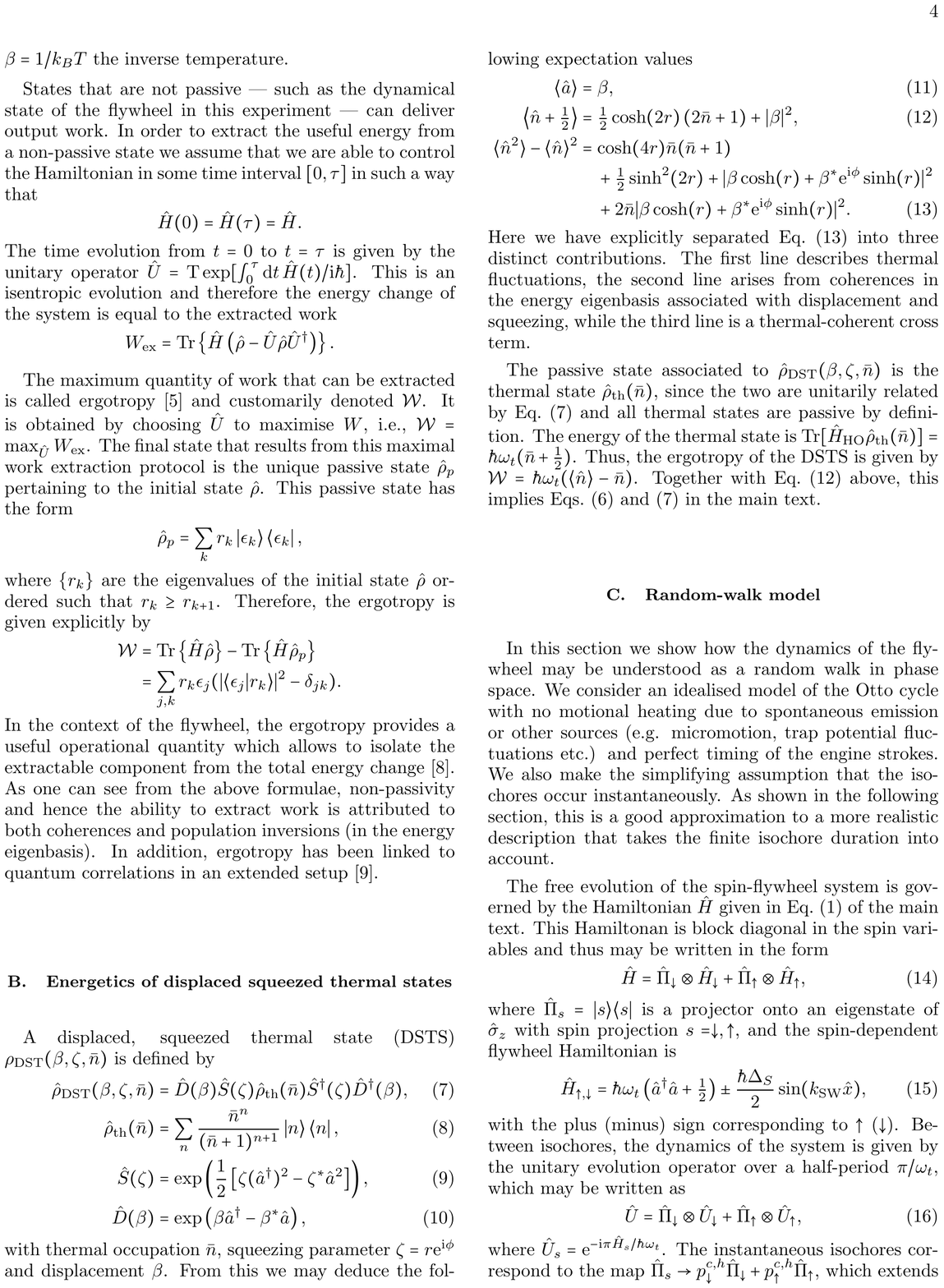}
\end{center}\end{figure}

\begin{figure}[htp]\begin{center}
\hspace*{-1cm}
\includegraphics[width=1.1\textwidth]{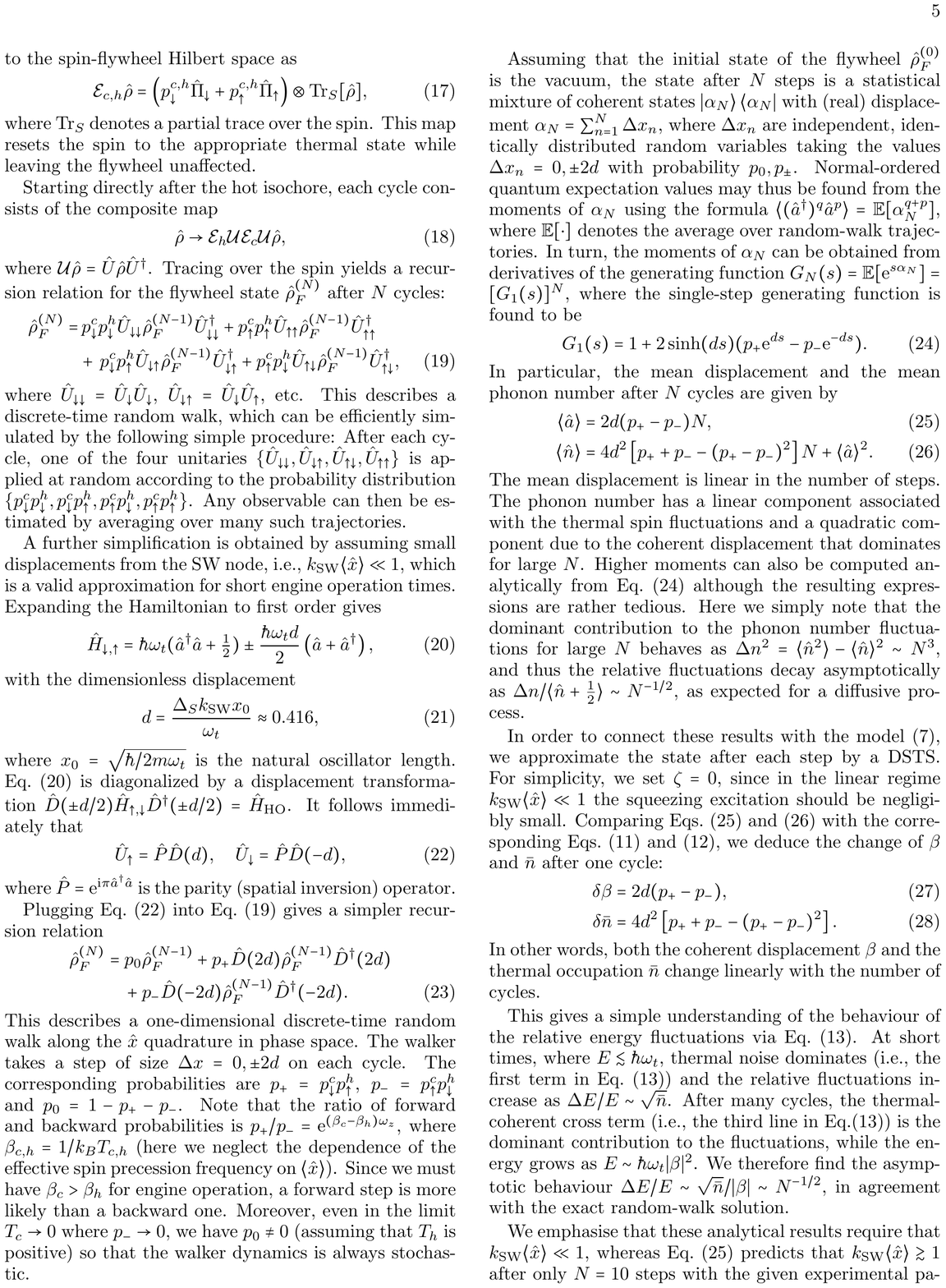}
\end{center}\end{figure}

\begin{figure}[htp]\begin{center}
\hspace*{-1cm}
\includegraphics[width=1.1\textwidth]{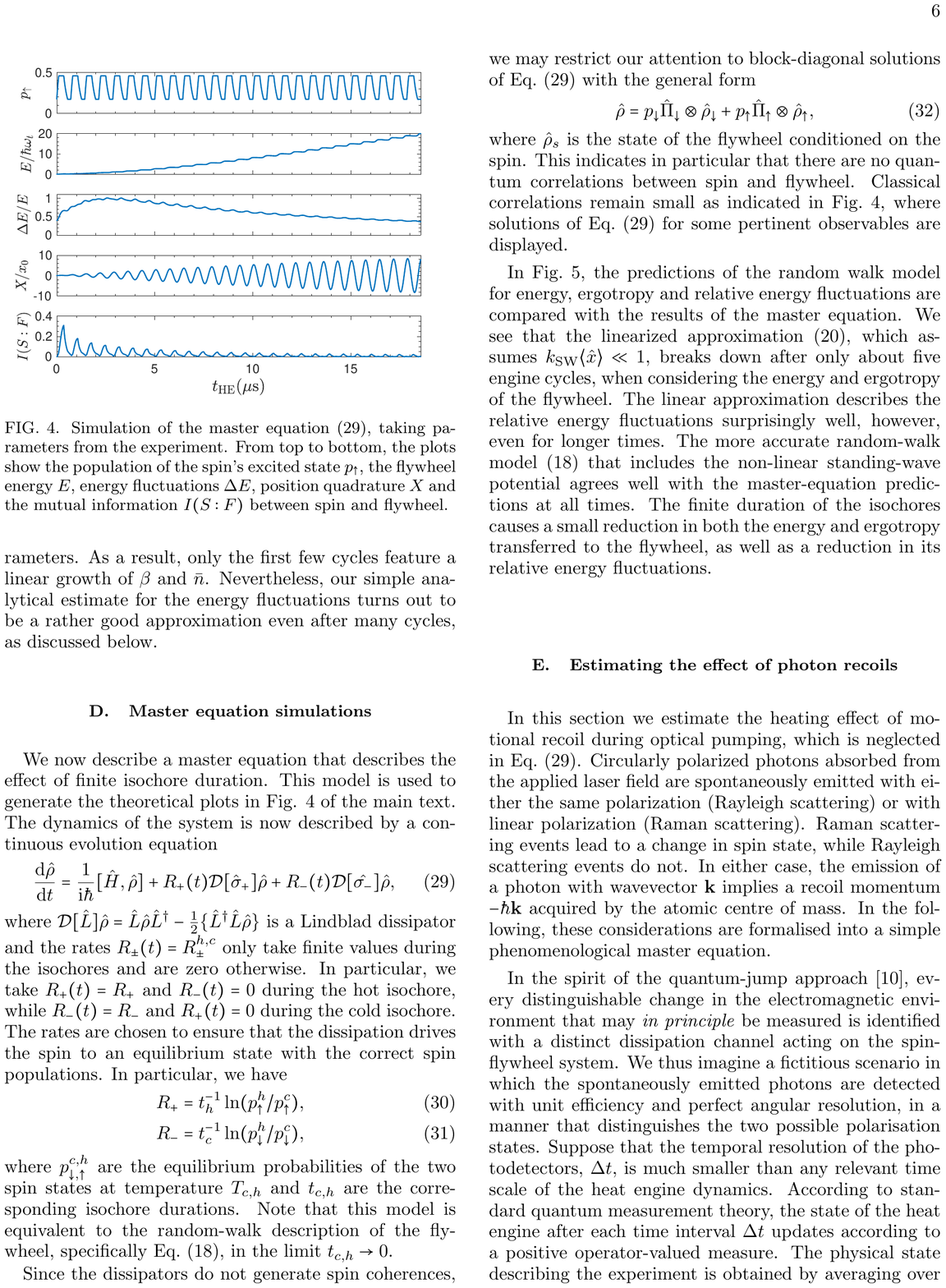}
\end{center}\end{figure}

\begin{figure}[htp]\begin{center}
\hspace*{-1cm}
\includegraphics[width=1.1\textwidth]{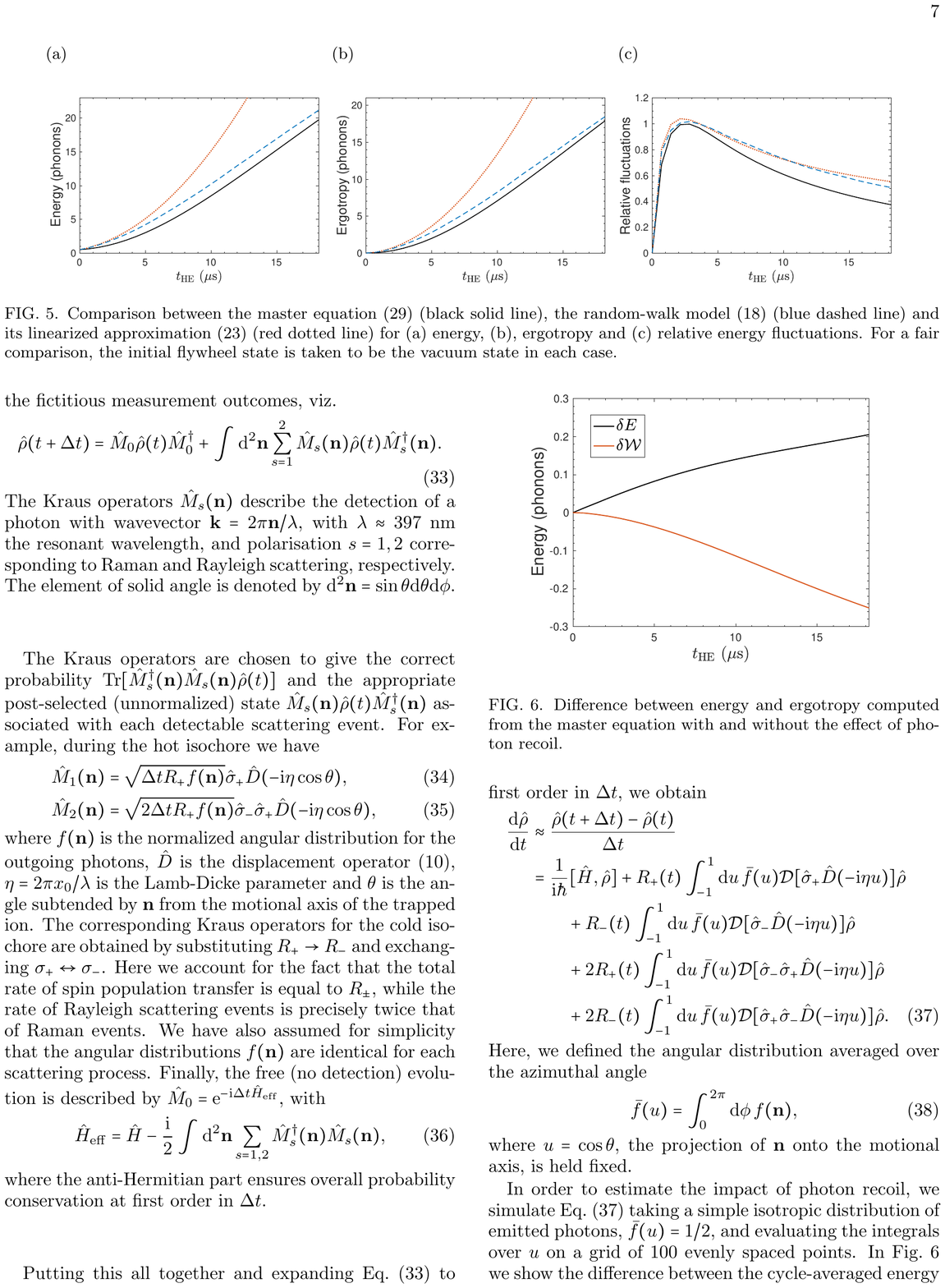}
\end{center}\end{figure}

\begin{figure}[htp]\begin{center}
\hspace*{-1cm}
\includegraphics[width=1.1\textwidth]{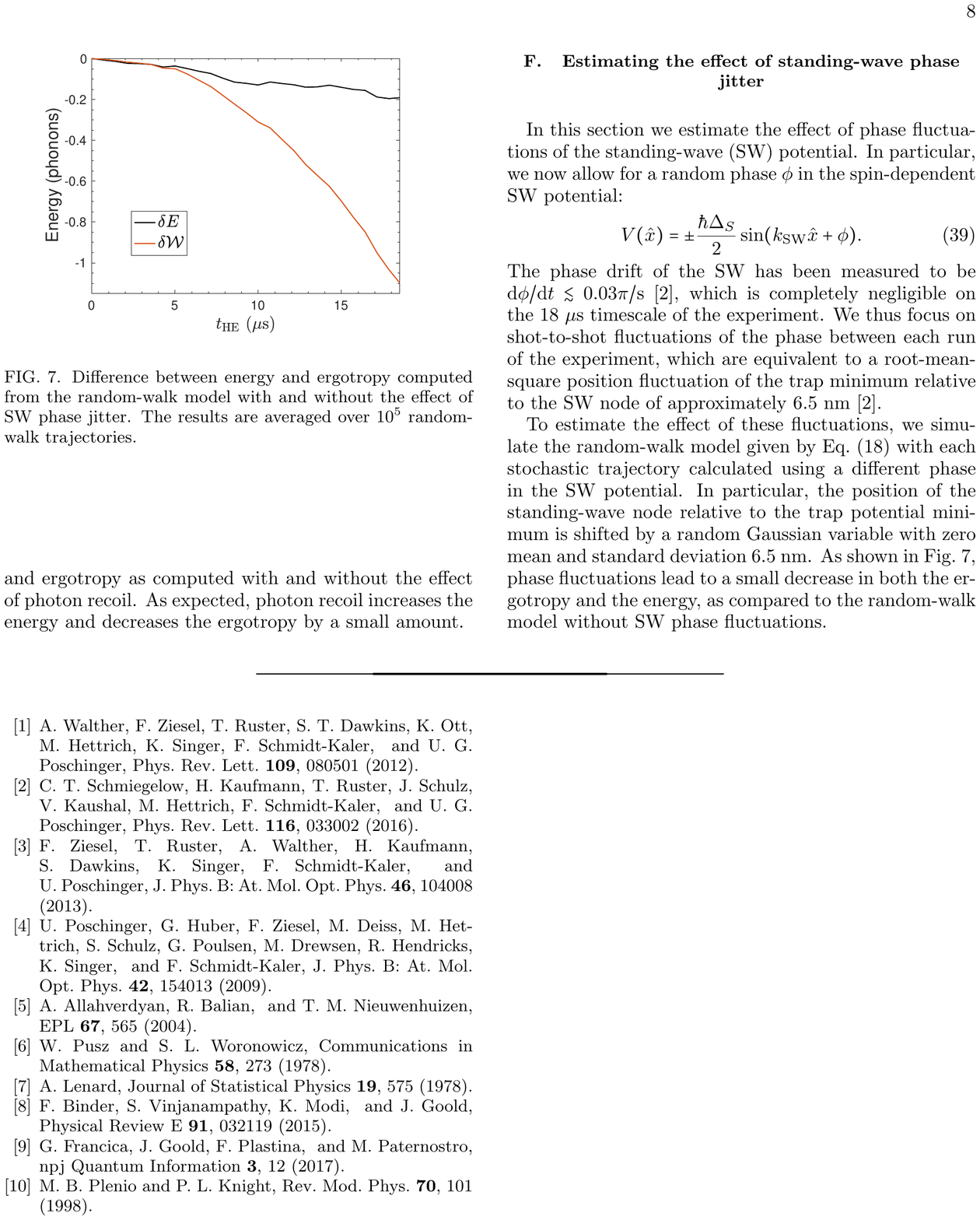}
\end{center}\end{figure}

\end{document}